\documentclass[usenatbib,usegraphicx]{mn2e}
\usepackage{amssymb,bm,color,multirow}

\pdfoutput=1

\newcommand{\e}[1]{\times 10^{#1}}

\title[Accretion to Stars with Complex Magnetic Fields]{Three-dimensional Simulations of Accretion to Stars with Complex Magnetic Fields}

\author[M. Long et al.]
{M. Long,$^1$\thanks{E-mail:long@astro.cornell.edu}, M.M. Romanova,$^1$\thanks{romanova@astro.cornell.edu}, and R.V.E. Lovelace, $^{1,2}$\thanks{lovelace@astro.cornell.edu}\\
$^1$ Department of Astronomy, Cornell University, Ithaca, NY 14853-6801, USA\\
$^2$ Department of Applied and Engineering Physics, Cornell University, Ithaca, NY 14853-6801, USA}

\begin{document}

%\date{Accepted . Received ; in original form }

%\pagerange{\pageref{firstpage}--\pageref{lastpage}} \pubyear{2006}

\maketitle

%\label{firstpage}

\begin{abstract}

Disk accretion to rotating stars with complex magnetic fields is investigated using full three-dimensional
magnetohydrodynamic (MHD) simulations. The studied magnetic configurations include superpositions of misaligned dipole
and quadrupole fields and off-centre dipoles. The simulations show that when the quadrupole component is comparable
to the dipole component, the magnetic field has a complex structure with three major magnetic poles on the surface of
the star and three sets of loops of field lines connecting them. A significant amount of matter flows to the quadrupole
``belt", forming a ring-like hot spot on the star. If the maximum strength of the magnetic field on the star is fixed,
then we observe that the mass accretion rate, the torque on the star, and the area covered by hot spots are several
times smaller in the quadrupole-dominant cases than in the pure dipole cases. The influence of the quadrupole component on
the shape of the hot spots becomes noticeable when the ratio of the quadrupole and dipole field strengths
$B_q/B_d\gtrsim0.5$. It becomes dominant in determining the shape of the hot spots when $B_q/B_d\gtrsim 1$. We conclude
that if the quadrupole component is larger than the dipole one, then the shape of the hot spots is determined by the quadrupole
field component. In the case of an off-centre dipole field, most of the matter flows through a one-armed accretion stream,
forming a large hot spot on the surface, with a second much smaller secondary spot. The light curves may have simple,
sinusoidal shapes, thus mimicking stars with pure dipole fields. Or, they may be complex and unusual. In some cases the
light curves may be indicators of a complex field, in particular if the inclination angle is known independently. We
also note that in the case of complex fields, magnetospheric gaps are often not empty, and this may be important for the
survival of close-in exosolar planets.

\end{abstract}


\section{Introduction}

The magnetic field of a rotating star can have a strong influence on the matter in an accretion disk. The field can disrupt
the disk and channel the accreting matter to the star along the field lines. The associated magnetic activity can be
observed through photometric and spectral measurements of different types of stars, like young solar-type
Classical T Tauri stars (CTTSs) (Hartmann et al. 1994), X-ray pulsars and millisecond pulsars (Ghosh \& Lamb 1978;
Chakrabarty et al. 2003), cataclysmic variables (Wickramasinghe et al. 1991; Warner 1995, 2000), and also brown dwarfs
(e.g., Scholz \& Ray 2006).  The topology of the stellar magnetic field plays an important role in disk accretion
and in the photometric and spectral appearance of the stars.

The  dipole magnetic model has been studied since Ghosh \& Lamb (1979a,b), both theoretically and with 2D and 3D MHD
simulations. However, the intrinsic field of the star may be more complex than a dipole field.  Safier (1998) argued
that the magnetic field of CTTSs may be strongly non-dipolar.  Zeeman measurements of the magnetic field of CTTSs based
on photospheric spectral lines show that the strong ($1-3$ kG) magnetic field is probably not ordered, and indicates
that the field is non-dipolar close to the star (Johns-Krull et al. 1999, 2007). Other magnetic
field measurements with the Zeeman-Doppler imaging technique have shown that in a number of rapidly rotating low-mass stars, the
magnetic field has a complicated multipolar topology close to the star (Donati \& Cameron 1997; Donati et al. 1999;
Jardine et al. 2002). Recently, observations from the ESPaDOnS/NAVAL spectropolarimeter (Donati et al. 2007a) revealed that
the magnetic field geometry of CTTSs BP Tau, V2129 Oph and SU Aur is more complex than dipole.

Theoretical and numerical research has been done on accretion to a rotating star with a non-dipole field. Jardine
et al. (2006) investigated the possible paths of the accreting matter in the case of a multipolar field derived from
observations. Gregory et al. (2006) developed a simplified stationary model for such accretion. von Rekowski and
Brandenburg (2006) did axisymmetric simulations, and von Rekowski and Piskunov (2006) did 3D simulations of the
disk-magnetosphere interaction in the case where the magnetic field is generated by dynamo processes in both the star and
the disk. They obtained a time-variable magnetic field of the star with a complicated multipolar configuration which
shows that the dynamo may give rise to a complex magnetic field structure. Recent observations and analysis of the
magnetic field and brightness distribution on the surface of CTTSs suggest that the magnetic field is predominantly
octupolar with strength $B_{oct}\sim 1.2$ kG and a much smaller dipole component $B_d\sim0.35$ kG (Donati et al. 2007b).
Most of the accretion luminosity is concentrated in a large spot close to the magnetic pole. Such observations are
interesting and can be compared with different 3D MHD simulation models. Here we investigate, using 3D MHD simulations,
disk accretion to rotating stars with dipole plus \textit{quadrupole} fields and off-centre dipole fields.

In our earlier work we performed full 3D simulations of disk accretion to stars with pure quadrupole and
\textit{aligned} dipole plus quadrupole fields (Long, et al. 2007) and found that the funnel streams and associated hot
spots have different features compared with the case of accretion to a star with pure dipole field. When the quadrupole component is significant, matter always accretes to a quadrupole ``belt" and forms a ring-like hot
spot on the surface of the star.

In this paper we show the results of full 3D MHD simulations of accretion to stars with more complex magnetic fields.
Compared with Long et al. (2007), we consider the more general case where the dipole and quadrupole are
\textit{misaligned} (see \S 3). Also, we consider the case of accretion to a star with internal
point dipoles displaced from the star's center (see \S 3.4).  Further, we performed a set of simulations with the aim
of understanding the properties of stars with different contributions from the dipole and quadrupole components.

One point of interest is to determine the how the area of the star's surface covered by hot spots depends on the the
quadrupole/dipole ratio. A further point of interest is to determine how the torque on the star depends on the
structure of the field. That is, we can compare the torque for cases with a complex field with those having a dipole
field. We discuss all these points in \S 4. We calculate the light curves and discuss the use of the light curves to
investigate the structure of the magnetic field of the star. The conclusions and discussion of possible applications of
our results are given in \S 5. Some high resolution figures and animations are available at
http://www.astro.cornell.edu/$\sim$long.

\section{The numerical model and the magnetic configurations}

\subsection{Model}

In a series of previous papers (Koldoba et al. 2002; Romanova et al. 2004; Ustyugova et al. 2006, Long, et al.
2005, 2007), we described the model used in our 3D MHD simulations. The model used in this paper is similar to that of
Long et al. (2007) and therefore it is only briefly summarized here.

We consider a rotating magnetized star surrounded by an accretion disk with a low-density, high-temperature corona
above and below the disk. The disk and the corona are initially in a quasi-equilibrium state. We solve the 3D MHD
equations in a reference frame rotating with the star, with the $z-$axis aligned with the star's rotation axis. The
magnetic field is decomposed into a ``main" component $\bm{B_0}$ and a variable component $\bm{B_1}$.  Here, $\bm{B_0}$
is time-independent in the rotating reference frame and consists of the dipole and quadrupole parts:
$\bm{B}_0=\bm{B}_d+\bm{B}_q$. Further, $\bm{B_1}$ is in general time dependent due to  currents in the simulation
region and is calculated from the MHD equations in the rotating reference frame.

A special ``cubed" sphere grid with the advantages of both spherical and cartesian coordinate systems was developed
(Koldoba et al. 2002). It consists of $N_r$ concentric spheres, where each sphere represents an inflated cube. Each
sphere consists of 6 sectors corresponding to the 6 sides of the cube, and an $N\times N$ grid of curvilinear cartesian coordinates is introduced in each sector. Thus, the whole simulation region consists of six blocks with $N_r\times N^2$ cells. In the current simulations we chose a grid resolution of $75\times31^2$. Other resolutions were also investigated for comparison. The coarser grids give satisfactory results for a pure dipole field. However, for a quadrupole field, the code requires a finer grid because of the higher magnetic field gradients. The code is a second order Godunov-type numerical scheme (see, e.g., Toro 1999)  developed earlier in our group (Koldoba et al. 2002). Viscosity is incorporated into the MHD equations in the interior of the disk so as to control the rate of matter inflow to the star. For the viscosity we used an $\alpha-$prescription with $\alpha=0.04$ in all simulation runs, and with smaller/larger values for testing.

\smallskip

\noindent\textbf{Initial Conditions:} The region considered consists of the star located in the center of coordinate
system, a dense disk located in the equatorial plane and a low-density corona which occupies the rest of the simulation
region. Initially, the disk and corona are in {\it rotational} hydrodynamic equilibrium. That is, the sum of the
gravitational, centrifugal, and pressure gradient forces is zero at each point of the simulation region. The initial
magnetic field is a combination of dipole and quadrupole field components which are force-free at $t=0$. The initial
rotational velocity in the disk is close to, but not exactly, Keplerian (i.e., the pressure gradient is taken into account). The corona at different cylindrical radii $r$ rotates with angular velocities corresponding to the Keplerian velocity of the disk at this distance $r$. This initial rotation is assumed so as to avoid a strong initial discontinuity of the magnetic field at the boundary between the disk and corona. The distribution of density and pressure in the disk and corona and the complete description of these initial conditions is given in Romanova et al. (2002) and Ustyugova et al. (2006).

The initial accretion disk extends inward to an inner radius $r_d$ and has a temperature $T_d$ which is much less than
the corona temperature $T_d=0.01T_c$. The density of the corona is $100$ times less than the density of the disk,
$\rho_c=0.01\rho_d$. These values of $T_c,T_d,\rho_c,\rho_d$ are specified at the disk-corona boundary near
the inner radius of the disk.

\smallskip

\noindent\textbf{Boundary Conditions:} At the inner boundary ($r=R=R_*$), where $R_*$ is the radius of the star,
boundary conditions are applied to the density ${\partial\rho}/{\partial r}=0$, pressure  ${\partial p}/{\partial
r}=0$, entropy ${\partial S}/{\partial r}=0$, velocity ${\partial({\bf v}-{\bf \Omega}\times{\bf R})}/{\partial r}=0$
and the magnetic field, ${\partial B_\theta}/{\partial r}=0$, ${\partial B_\phi}/{\partial r}=0$. The $r-$component of
the magnetic field satisfies ${\partial (r^2B_r)}/{\partial r}=0$. The matter flow is frozen to the strong magnetic
field so that we have $(\bf{v}-\bf{\Omega}\times\bf{R})\parallel\bf{B}$. At the outer boundary, free boundary
conditions are taken for all variables with the additional condition that matter is not permitted to flow in through the
outer boundary. The investigated numerical region is large, $\sim 45 R_*$, so that the  initial reservoir of matter in
the disk is large and sufficient for the performed simulations. Matter flows slowly inward from the external regions of the disk. During the simulation times studied here only a small fraction of the total disk matter accretes to the star.

\subsection{Reference Units:}

We solve the MHD equations using dimensionless variables: distance $\widetilde{R}=R/R_0$, velocity
$\widetilde{v}=v/v_0$, time $\widetilde{t}=t/P_0$, etc. The subscript ``0" denotes a set of reference (dimensional)
values for variables, which are chosen as follows: $R_0=R_*/0.35$, where $R_*$ is the radius of the star;
$v_0=(GM/R_0)^{1/2}$; time scale $P_0=2\pi R_0/v_0$. Other reference values are: angular velocity
$\Omega_0=v_0/R_0$; magnetic field $B_0=B_{*0}(R_*/R_0)^3$, where $B_{*0}$ is the reference magnetic field on the
surface of the star; dipole magnetic moment $\mu_0=B_0R_0^3$; quadrupole moment $D_0=B_0R_0^4$; density
$\rho_0=B_0^2/v_0^2$; pressure $p_0=\rho_0v_0^2$; mass accretion rate $\dot{M}_0=\rho_0v_0R_0^2$; angular momentum flux
$\dot{L}_0=\rho_0v_0^2R_0^3$; energy per unit time $\dot{E}_0=\rho_0v_0^3R_0^2$ (the radiation flux $J$ is also in
units of $\dot{E}_0$, see \S3.1); temperature $T_0=\mathcal{R}p_0/\rho_0$, where $\mathcal{R}$ is the gas constant;
and the effective blackbody temperature $T_{\mathrm{eff,0}} = (\rho_0 v_0^3/\sigma)^{1/4}$, where $\sigma$ is the
Stefan-Boltzmann constant.

In the subsequent sections and figures, we show dimensionless values for all quantities and drop the tildes ($\sim$). To obtain the real dimensional values of variables, one needs to multiply the dimensionless values by the corresponding
reference units. Our dimensionless simulations are applicable to different astrophysical objects with different scales.
For convenience, we list the reference values for typical CTTSs, cataclysmic variables, and millisecond pulsars in Tab.
1.

\begin{table}
% \centering
% \newcommand\ts{\rule{0pt}{2.6ex}}
% \newcommand\bs{\rule[-1.2ex]{0pt}{0pt}}
\begin{tabular}{l@{\extracolsep{0.2em}}l@{}lll}

\hline
&                                                   & CTTSs       & White dwarfs          & Neutron stars           \\
\hline

\multicolumn{2}{l}{$M_*(M_\odot)$}                  & 0.8         & 1                     & 1.4                     \\
\multicolumn{2}{l}{$R_*$}                           & $2R_\odot$  & 5000 km               & 10 km                   \\
\multicolumn{2}{l}{$B_{*0}$ (G)}                    & $10^3$      & $10^6$                & $10^9$                  \\
\multicolumn{2}{l}{$R_0$ (cm)}                      & $4\e{11}$   & $1.4\e9$              & $2.9\e6$                \\
\multicolumn{2}{l}{$v_0$ (cm s$^{-1}$)}             & $1.6\e7$    & $3\e8$                & $8.1\e9$                \\
\multicolumn{2}{l}{$\Omega_0$ (s$^{-1}$)}           & $4\e{-5}$   & 0.2                   & $2.8\e3$                \\
\multicolumn{2}{l}{\multirow{2}{*}{$P_0$}}          & $1.5\e5$ s  & \multirow{2}{*}{29 s} & \multirow{2}{*}{2.2 ms} \\
&                                                   & $1.8$ days  &                       &                         \\
\multicolumn{2}{l}{$B_0$ (G)}                       & 43          & $4.3\e4$              & $4.3\e7$                \\
\multicolumn{2}{l}{$\rho_0$ (g cm$^{-3}$)}          & $7\e{-12}$  & $2\e{-8}$             & $2.8\e{-5}$             \\
\multicolumn{2}{l}{$p_0$ (dy cm$^{-2}$)}            & $1.8\e{3}$  & $1.8\e{9}$            & $1.8\e{15}$             \\
\multirow{2}{*}{$\dot M_0$} & (g s$^{-1}$)          & $1.8\e{19}$ & $1.2\e{19}$           & $1.9\e{18}$             \\
                            & ($M_\odot$yr$^{-1}$)  & $2.8\e{-7}$ & $1.9\e{-7}$           & $2.9\e{-8}$             \\
\multicolumn{2}{l}{$\dot{L}_0$ (g cm$^2$s$^{-2}$)}  & $1.15\e{18}$& $4.9\e{36}$           & $4.5\e{34}$             \\
\multicolumn{2}{l}{$T_0$ (K)}                       & $1.6\e6$    & $5.6\e8$              & $3.9\e{11}$             \\
\multicolumn{2}{l}{$\dot E_0$ (erg s$^{-1}$)}       & $4.8\e{33}$ & $1.2\e{36}$           & $1.2\e{38}$             \\
\multicolumn{2}{l}{$T_{\mathrm{eff},0}$ (K)}        & 4800        & $3.2\e5$              & $2.3\e7$                \\
%T_eff_0 for B_*0=10^8 is 7.2\e6
\hline
\end{tabular}
\caption{Sample reference units for typical CTTSs, cataclysmic variables, and millisecond pulsars. Real dimensional
values for variables can be obtained by multiplying the dimensionless values of variables shown in this paper by these reference units. See also \S2.2.} \label{tab:refval}
\end{table}

\subsection{Magnetic Configurations}

\noindent\textbf{Combination of Dipole and Quadrupole Fields:} The intrinsic magnetic field of the star is
$\mathbf{B}=-\nabla\varphi$, where the scalar potential of the magnetic field is $\varphi(\mathbf{r})=
\Sigma{m_a}/{|\mathbf{r-r}_a|}$, $m_a$ is the magnetic ``charge" analogous to the electric charge, and $\mathbf{r}$ and $\mathbf{r}_a$ are the positions of the observer and the magnetic ``charges" respectively. The scalar potential can be represented as a
multipole expansion in powers of $1/r$ and the quadrupole term is
\[
\varphi^ {(2)}={D_{\alpha\beta}n_{\alpha}n_{\beta}}/{2r^3},
\]
where $n_{\alpha}=x_\alpha/r$, $x_\alpha$ are the components of $\mathbf{r}$, $D_ {\alpha\beta}$ is the magnetic quadrupole moment tensor, summation over repeated indices is implied, and
$D_{\alpha\alpha}=0$. In the \textit{axisymmetric} case where $D_{11}=D_{22}=-D_{33}/2$, let $D=D_{33}$ be the value of
the quadrupole moment and refer to the axis of symmetry as the ``direction'' of the quadrupole moment, so we have the
quadrupole moment $\bm{D}$. The combination of dipole and quadrupole magnetic fields can be written as:
\begin{equation}
\mathbf{B(r)}=\frac{3(\bm{\mu}\cdot{\hat{\bf r}})\hat{\bf r}-\bm{\mu}}{r^3}+\frac{3D}{4r^4}(5(\hat{\bf D}\cdot\hat{\bf r})^2-1)\hat{\bf r}-\frac{3D}{2r^4}(\hat{\bf D}\cdot\hat{\bf r})\hat{\bf D},
\end{equation}
where $\hat{\bf r}$ and $\hat{\bf D}$ are the unit vectors for the position and the quadrupole moment respectively.
In general, the dipole and quadrupole moments $\bm{\mu}$ and $\bm{D}$ are misaligned relative to the
rotational axis $\bm{\Omega}$, at angles $\Theta$ and $\Theta_D$ respectively. In addition,
they can also be in different meridional planes with an angle $\Phi$ between the $\bm{\Omega-\mu}$ and $\bm{\Omega-D}$
planes.

The values of $\mu$ and $D$ determine the magnetic field strength on the star's surface. For example, if $\bm{\mu}$ and $\bm{D}$ are aligned with the rotational axis $\bm{\Omega}$, and $\mu=0.5$ , $D=0.5$, then the strength of the dipole component on the north pole of the star is $B_{*d}=0.68$ kG, and the quadrupole one is $B_{*q}=0.32$ kG.

\smallskip

\noindent\textbf{Off-centre Magnetic Configurations:} The magnetic field of a star may deviate from the center of
the star. Such displacement naturally follows from the dynamo mechanism of magnetic field generation. If the
dipole and quadrupole moments are displaced by $\bm{a}$ and $\bm{b}$ from the center of the star, we can replace
$\bm{r}$ with $\bm{(r-a)}$ and $\bm{(r-b)}$ in the corresponding terms in Eqn. 1 to obtain the magnetic field around the
star. In this paper we investigate accretion to a displaced dipole and to a set of displaced dipoles.

\section{Accretion to Stars with Different Complex Magnetic Fields Configurations}

In Long et al. (2007), we have shown results of accretion to stars with dipole plus quadrupole fields with aligned axes. In this section, we present simulation results of accretion to a star with \textit{misaligned} dipole plus
quadrupole magnetic fields. First, we consider the configuration when the dipole magnetic moment $\bm\mu$ is aligned
with the stellar spin axis $\bm\Omega$, that is, $\Theta=0^\circ$,  but the quadrupole magnetic moment $\bm{D}$ is
inclined relative to the spin axis at an angle $\Theta_D=45^\circ$. Next, we consider a more general configuration when
$\bm\mu$, $\bm{D}$ and $\bm\Omega$ are all misaligned relative to each other, and there is an angle $\Phi$ between
the $(\bm{\Omega-\mu})$ and $(\bm{\Omega-D})$ planes. To ensure that the case is general enough to represent a possible situation in nature, we choose $\Theta=30^\circ$, $\Theta_D=45^\circ$ and $\Phi=90^\circ$.

The strengths of the dipole and quadrupole moments are the same in both cases, $\mu=D=0.5$. Here we choose a quite
strong quadrupole component to get an example of a complex magnetic field, so that the disk is disrupted not by the
dipole component only but by the superposition of the dipole and quadrupole fields. Fig. 1 shows the distribution of
magnetic fields for these two cases in $+x$ direction. We can see that the quadrupole component is still comparable to
the dipole component at $r\sim1$ where the disk is stopped by the magnetosphere (see Fig. 5). This means that both dipole and quadrupole contribute to stop the disk.

\begin{figure}
\begin{center}
\includegraphics{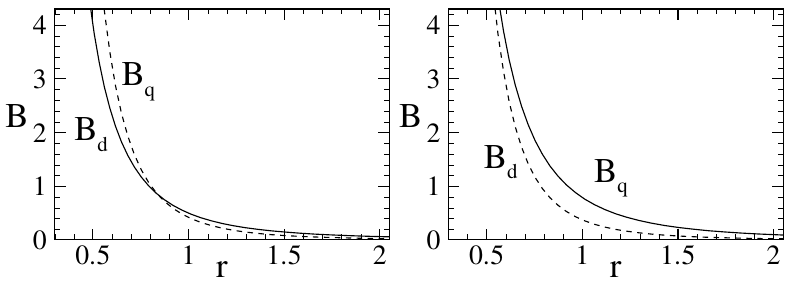}
\caption{\label{fig1}The distribution of dipole and quadrupole components of magnetic fields. The left-hand panel
represents the case when $\mu=D=0.5$, $\Theta=0^\circ$,$\Theta_D=45^\circ$; the right-hand panel represents the case
when $\mu=D=0.5$, $\Theta=30^\circ$, $\Theta_D=45^\circ$, $\Phi=90^\circ$. The strengths of the quadrupole and dipole are comparable at $r\sim1$ where the disk is stopped by the magnetosphere.}
\end{center}
\end{figure}

\subsection{Misaligned dipole plus quadrupole configurations}

Now we consider the configuration when the $(\bm{\Omega-\mu})$ and $(\bm{\Omega-D})$ planes coincide: $\mu=D=0.5$,
$\Theta=0^\circ$, $\Theta_D=45^\circ$. Fig. 2 shows the strength of the magnetic field on the surface of the star in
different projections. One can see that the field is not symmetric and shows three strong magnetic poles and a much
weaker magnetic pole with different polarities on the surface, due to the asymmetry of the misaligned dipole plus
quadrupole configuration. The left-hand panel shows a strong positive magnetic pole (red) with $B=67$ (dimensionless
value) and a strong negative pole (dark blue) with $B=-41$. The middle panel shows a second strong positive pole
(yellow) with $B=35$, and in fact a very weak and small negative pole in the middle of the surface with $B=-7$ (dark
green). We should note that the strength of the magnetic field around this weak pole is more negative than at the pole. The right-hand panel shows both the strong negative pole (dark blue) which is extended and the second strong positive pole (yellow).

\begin{figure}
\begin{center}
\includegraphics{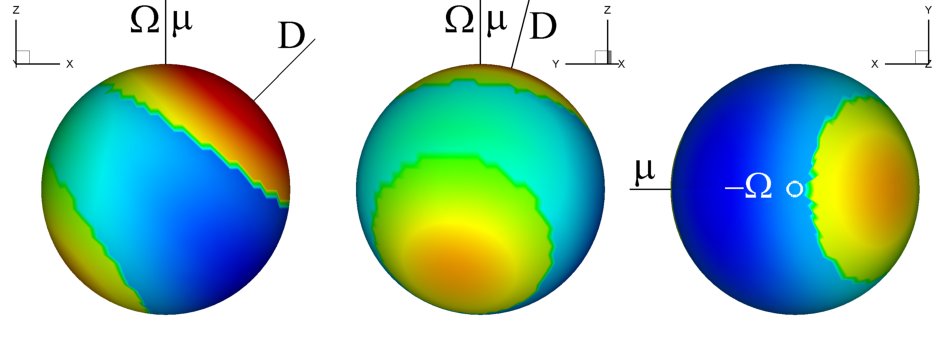}
\caption{\label{fig2} The strength of the magnetic field on the surface of the star for the case $\mu=0.5$, $D=0.5$,
$\Theta=0^\circ$, $\Theta_D=45^\circ$ at different projections: from the equatorial plane (left-hand panel), close to the $yz$ plane (middle panel), and from the south pole (right-hand panel). The red and blue regions represent different strength and polarities. The strength varies from the most blue point $B=-41$ to the most red point $B=67$ (dimensionless value, see \S2.2 and Tab. 1) and reaches $B=0$ at the boundary between the yellow and green areas.}
\end{center}
\end{figure}

\begin{figure}
\begin{center}
\includegraphics{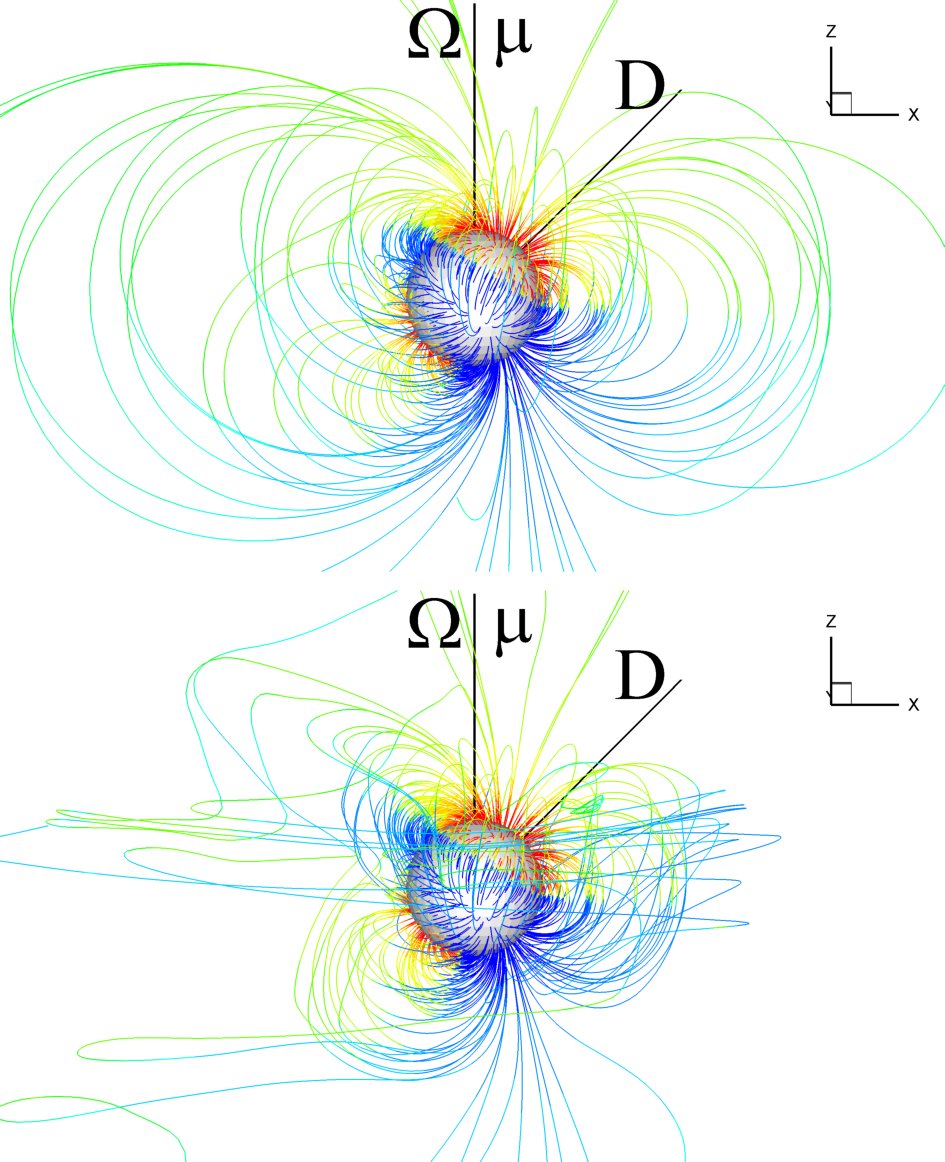}
\caption{\label{fig3} The complex magnetic field lines for the case $\mu=0.5$, $D=0.5$, $\Theta=0^\circ$,
$\Theta_D=45^\circ$, at $t=0$ (top panel) and $t=8$ (bottom panel). The strength of the magnetic field along the field
lines varies from red (positive magnetic pole) to blue (negative pole). }
\end{center}
\end{figure}

\begin{figure}
\begin{center}
\includegraphics{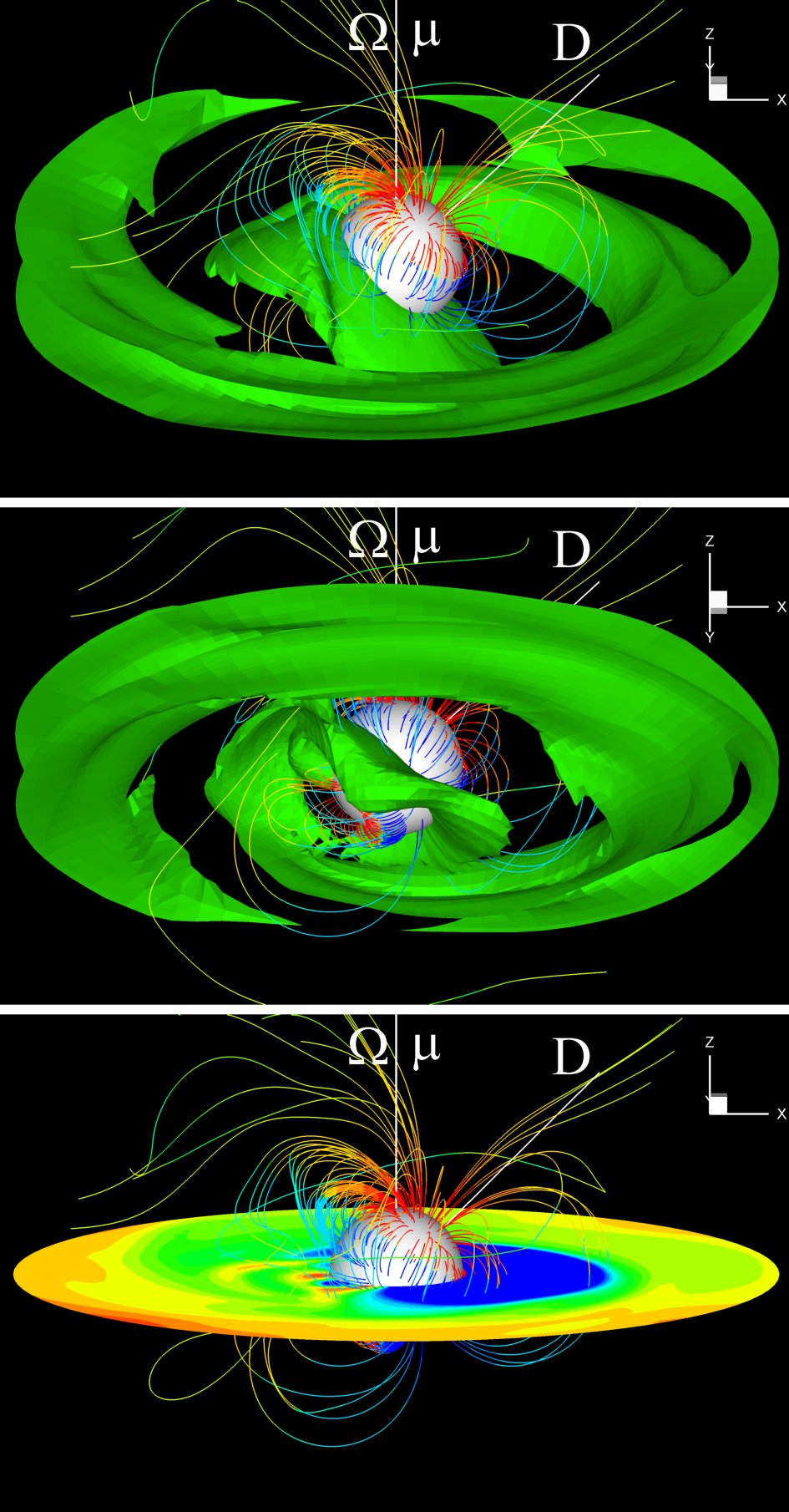}
\caption{\label{fig4} 3D views of matter flow to the star for the case $\mu=0.5$, $D=0.5$, $\Theta=0^\circ$,
$\Theta_D=45^\circ$, at $t=8$. The top panel shows the view from the north, the middle panel from the south.
The green surface is a constant density surface, $\rho=0.27$ (dimensionless value, see \S2.2). The bottom panel shows the equatorial plane. The colors of the magnetic field lines correspond to those in Fig. 3. Only the inner part of the simulation region is shown.}
\end{center}
\end{figure}

Fig. 3 shows a three-dimensional (3D) view of the structure of the magnetic field lines. The color shows the different
strengths and polarities of the magnetic field lines, from maximum positive (red) to maximum negative (blue) values.
The magnetic field lines are more complex than in the aligned dipole plus quadrupole case. Now we see three sets of loops
of closed field lines in the region near the star, connecting the three major poles and the weaker pole on the star,
which is different from the one big and one small loops of field lines shown in the aligned dipole plus quadrupole case
(Long et al. 2007).

Fig. 4 shows a 3D view of matter flow around the star. The disk matter is disrupted by the complex field and is lifted
above the equatorial plane at the magnetospheric radius $r=r_m$, where the magnetic stress balances the matter stress,
$\beta=(p+\rho v^2)/(B^2/8\pi)=1$.  We can see that most of the matter flows between the loops of field lines to the extended negative magnetic pole by choosing the shortest path which is energetically favorable, forming a modified quadrupole ``belt". This ``belt" has a different shape from that in the case of aligned dipole plus quadrupole configurations (see Long et al. 2007), which looks like a more ``regular" sheet perpendicular to the magnetic axis. Here, the ``belt" is twisted, although it is approximately perpendicular to the quadrupole axis. The bottom panel shows a slice in the equatorial plane, and we can see that in the $-x$ direction, the matter penetrates between field lines to the surface of the star.

\begin{figure*}
\begin{center}
\includegraphics{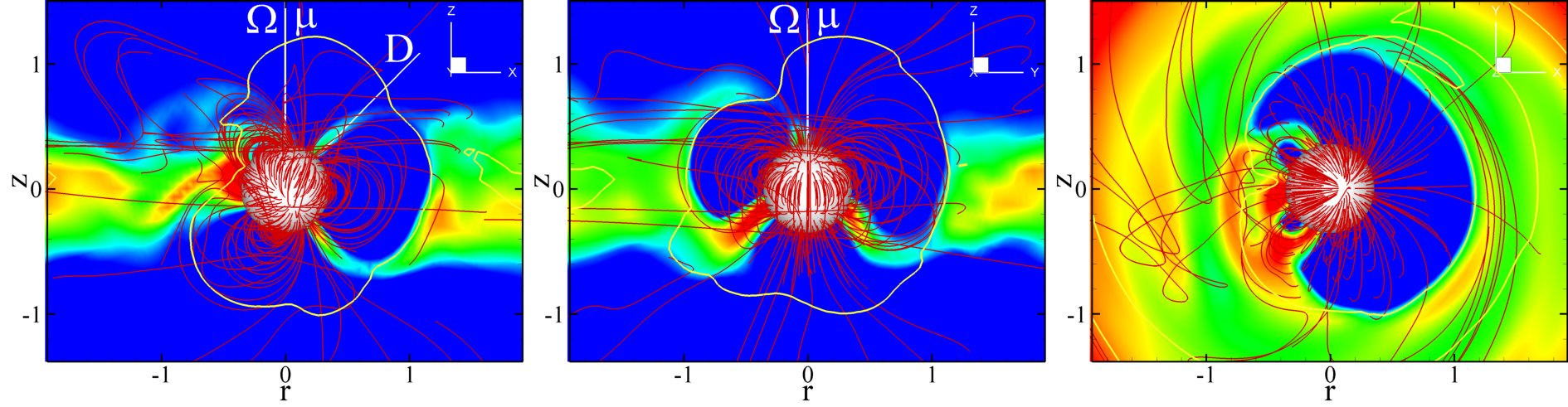}
\caption{\label{fig5} Disk accretion viewed from different projections for the case of $\mu=0.5$, $D=0.5$,
$\Theta=0^\circ$, $\Theta_D=45^\circ$. The left-hand panel shows the projection in $xz$ plane, the middle panel in the $yz$ plane, and the right-hand panel in the $xy$ plane. The color background shows the density distribution varying from $\rho=0.01$ (blue) to $\rho=1.9$ (red), and the magnetic field lines are shown in red. The yellow lines correspond to $\beta=1$.}
\end{center}
\end{figure*}

\begin{figure}
\begin{center}
\includegraphics{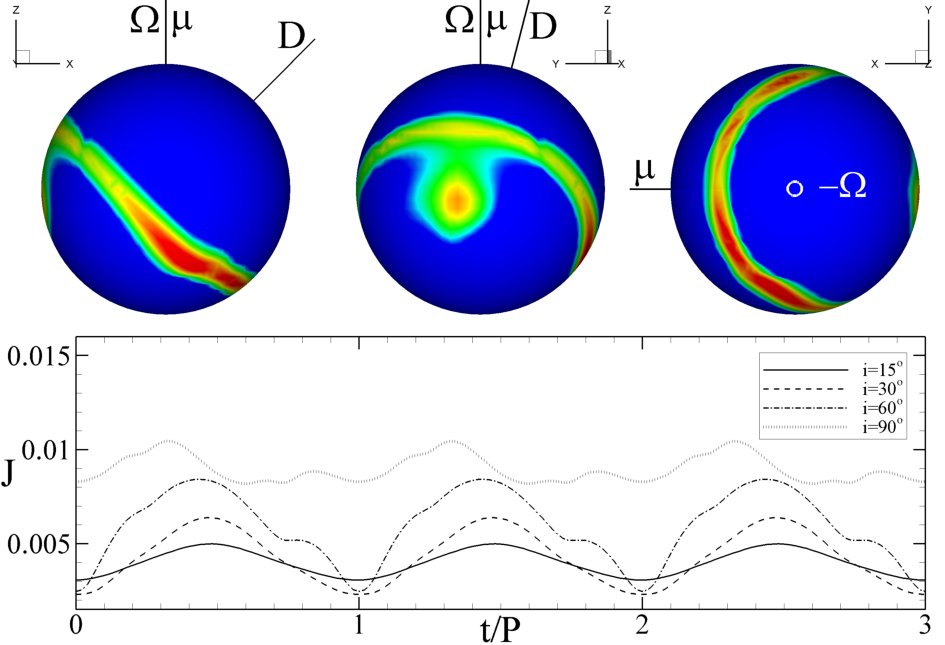}
\caption{\label{fig5} The hot spots viewed at different angles: from the rotational equatorial plane (left-hand panel);
from near the $yz$ plane (middle panel); and from the south pole (right-hand panel). The red color corresponds to the densest region with maximum density $\rho=1.4$ (dimensionless value, see \S2.2). The bottom panel shows the light curves for different inclination angles.}
\end{center}
\end{figure}

Fig. 5 shows different projections of the accretion flow. One can see that the matter flow is not symmetric in the $xz$ and $xy$ planes. The $xz$ projection shows that in $-x$ direction, the magnetic stress is weak due to the combination of the dipole and tilted quadrupole components, so that the matter is not lifted, and flows almost directly to the surface of the star, and the equatorial plane does not have the magnetospheric gap which is typical for pure dipole cases. Because the quadrupole moment is inclined at $45^\circ$ relative to the $z-$ axis in $xz$ plane, in the $+x$ direction, there is one set of closed loops of magnetic field lines, which stops the disk and forms a magnetospheric gap. Comparison of the dipole and quadrupole components in the $+x$ direction shows that in the region where the disk stops ($\beta=1$), the dipole and quadrupole components are approximately equal, so that the quadrupole has a strong influence on the matter flow, which is clearly seen from the above figures. In the $yz$ projection, the matter flow is relatively symmetric because $\bm{\Omega}$, $\bm\mu$ and $\bm{D}$ are all in the $xz$ plane. In the $xy$ projection, we can see that there is a gap in $+x$ direction where there is no direct matter flow, but there is no gap in the $-x$ direction.

Fig. 6 shows the hot spots on the surface of the star at $t=8$. One can see that there is a ring-like hot spot and a
small round hot spot nearby. There is no clear hot spot near the positive part of the $\bm{D}$ and $\bm{\Omega}$ axes,
due to the weak funnel stream in the northern hemisphere. The ring-like hot spot comes from accretion to the quadrupole
``belt", which is approximately symmetric relative to $\bm D$ because the quadrupole magnetic component $B_q$ is
stronger than the dipole component $B_d$ in the region close to the star. A small round spot forms as a
result of the connection of field lines between the weak magnetic pole ($B=-7$) and the nearby region where $B$ is
positive. Part of the matter flows to the south and forms this round hot spot.

We calculated the light curves from the hot spots on the rotating star's surface.  Assuming the total energy of the
inflowing matter is radiated isotropically as blackbody radiation, we obtain the flux of the radiation in a direction
$\bm{\hat{k}}$,
\begin{equation}
J=r^2F_{obs}=\int I(\bm{R,\hat{k}})\cos\theta\mathrm{d}S,
\end{equation}
where $r$ is the distance between the star and the observer, $F_{obs}$ is the observed flux, $I(\bm{R,\hat{k}})$ is
the specific intensity of the radiation from a position $\bm{R}$ on the star's surface into a solid angle element $\mathrm{d}\Omega$ in the direction $\bm{\hat{k}}$, $\theta=\arccos{(\bm{R}\cdot{\hat{k}})}$, $\mathrm{d}S$ is an element of the surface area. The specific intensity can be obtained as
\begin{equation}
I(\bm{R,\hat{k}})=\frac{1}{\pi}F_e(\bm{R})\cos\theta,
\end{equation}
where $F_e(\bm{R})$ is the total energy flux of the inflowing matter. In our simulations, $J$ is the received energy
per unit time, and the dimensionless value of $J$ is in units of $\dot{E}_0$ which is shown in Tab. 1 and discussed in
\S2.2. The hot spots constantly change their shape and location. However, our 3D simulations show that the changes are
relatively small. So we choose the hot spots at some moment of time, fix them, and rotate the star to obtain the light
curves. The bottom panel of Fig. 6 shows the light curves at $t=8$ for different inclination angles
$i=\arccos{(\hat{\bm\Omega}\cdot\bm{\hat{k}})}$. One can see that the light curve is approximately sinusoidal for small
inclination angles. This is because the observer can only see a part of the ring-like hot spots at this time and the
sinusoidal shape is determined by the rotation of the spots with the star. One can see that the shapes of the light
curves at $i=60^\circ$ and $90^\circ$ are unusual and do not correspond to any pure dipole field case (Romanova et al.
2004). When $i$ increases, the small round hot spot and more of the ring-like hot spots can be observed, and
consequently the peak intensity and variability of the light curves are larger.

\subsection{A more general case: $\bm{\mu,D,\Omega}$ all misaligned}

Next we investigate the more general case when the dipole moment $\bm\mu$, quadrupole moment $\bm{D}$ and spin axis $\bm{\Omega}$ of the star are all misaligned relative to each other: $\mu=0.5$, $D=0.5$, $\Theta=45^\circ$,
$\Theta_D=30^\circ$ and $\Phi=90^\circ$. Fig. 7 shows the distribution of the magnetic field on the surface of the
star. We can again see that there are three dominant magnetic poles with magnetic strengths $B=65, B=37 and B=-44$
respectively. There is another weak magnetic pole in the middle region shown in the middle panel, but it shrinks and
partially merges into the boundary of the nearby pole and looks smaller than that in the previous case. Fig. 8 shows 3D
plots of matter flow at $t=8$. In the top panel, one can see that again, a significant amount of matter flows through the quadrupole ``belt" below the loops of field lines, because the quadrupole component dominates near the star. And a single strong stream flows to the region nearby the north pole. So it could be expected that ring-like hot spots and round hot spots form on the surface of the star. The bottom panel shows a slice in the $xy$ plane. One can see that some matter can flow to the star directly without leaving the equatorial plane. We also can see three sets of loops of field lines.

Fig. 9 shows different projections of the matter flow. The matter flow is no longer symmetric in any projection which
is expected due to the magnetic configuration. In the $xy$ and $yz$ planes, we also can see that there are three sets of
loops of field lines which regulate the shape of the accretion flow. In the northern hemisphere, some matter is lifted from the equatorial plane and flows to the high-latitude area near the dipole axis. This is because the accretion disk is close to one of the major magnetic poles in this direction. In the southern hemisphere, matter penetrates the $\beta$ line between loops of field lines, and forms a modified quadrupole ``belt". The magnetosphere is not empty in the $xy$ plane, and matter can go directly to the star in some directions.

Fig. 10 shows the hot spots and the associated light curves. We can see there is one strong arc-like hot spot near the
dipole magnetic axis, but not at the magnetic pole. This hot spot comes from the one-armed stream shown in Fig. 9.
Another ring-like hot spot is below the equatorial plane but not symmetric with respect to either the
$\bm\Omega$, $\bm\mu$ or $\bm{D}$ axis. The light curve is sinusoidal for the small inclination angle $i=15^\circ$. At larger $i$ beside the round north spot, part of the ring-like spot becomes visible to the observer, so the light curves depart from the sinusoidal shapes and have a larger amplitude.

\begin{figure}
\begin{center}
\includegraphics{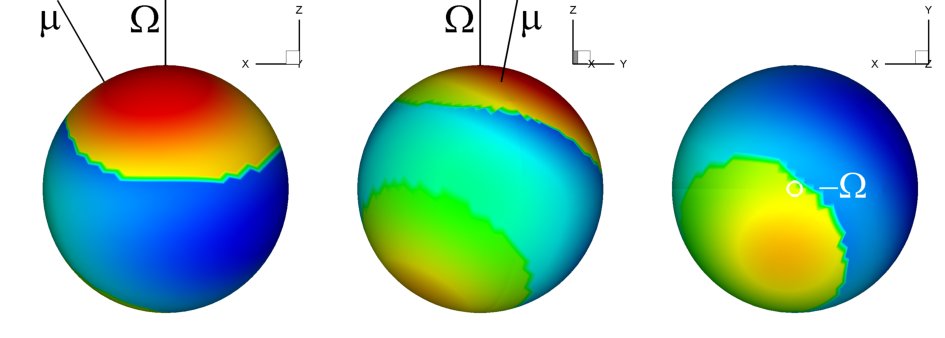}
\caption{\label{fig7} Distribution of the magnetic field strength on the star's surface for the case
$\mu=0.5$, $D=0.5$, $\Theta=45^\circ$, $\Theta_D=30^\circ$ and $\Phi=90^\circ$ at different projections: from
the equatorial plane (left-hand panel), from near the $yz$ plane (middle panel), and from the south pole (right-hand panel). Red and blue colors show positive and negative polarities.  The magnetic strength varies from $B=-44$ to $B=65$.}
\end{center}
\end{figure}

\begin{figure}
\begin{center}
\includegraphics{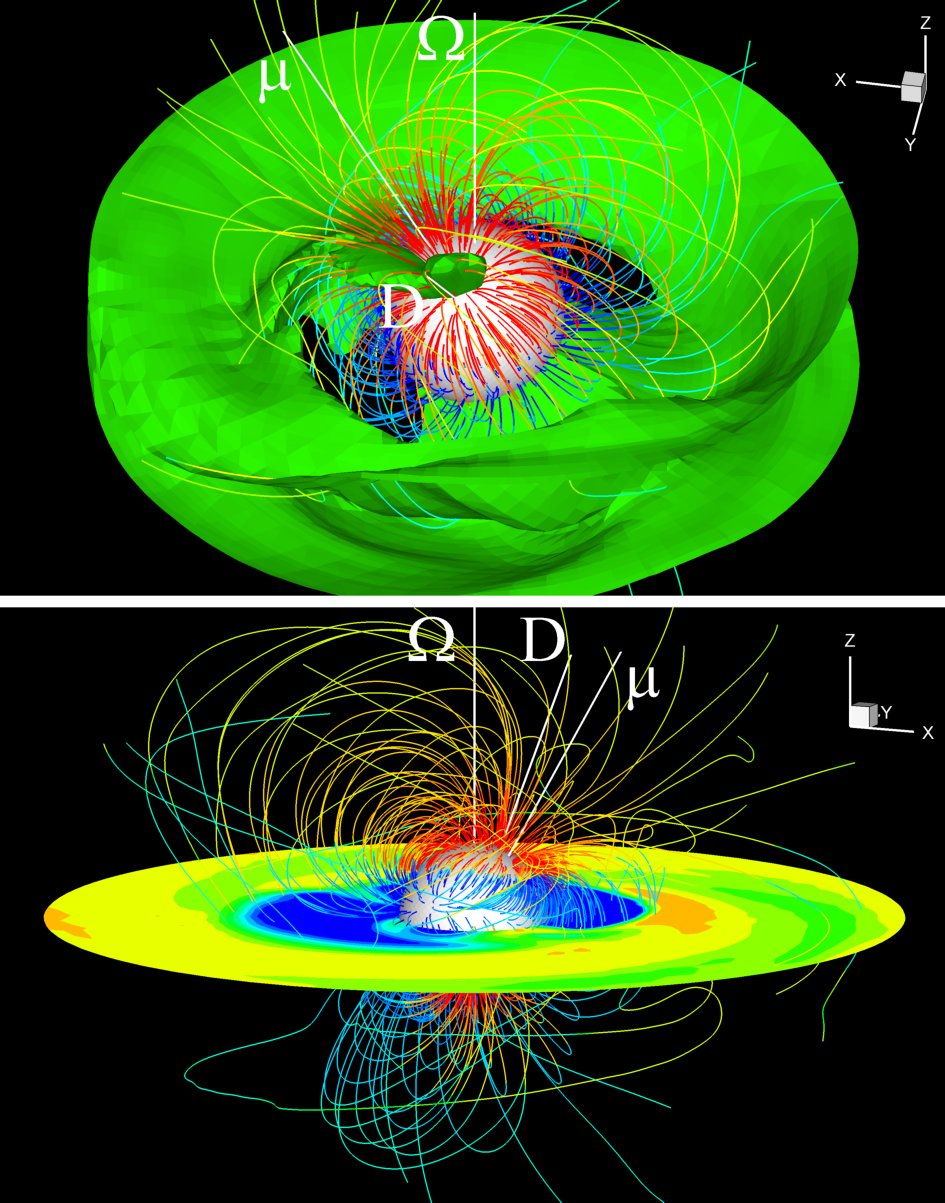}
\caption{\label{fig8} 3D views of disk accretion to a star in the more general misaligned dipole plus quadrupole
case at $t=8$, when $\mu=0.5$, $D=0.5$, $\Theta=45^\circ$, $\Theta_D=30^\circ$ and $\Phi=90^\circ$. A constant density
level ($\rho=0.25$, dimensionless value) of the disk is shown in green in the top panel; different density levels
in the equatorial plane are shown in the bottom panel. The field lines are shown in multiple colors. }
\end{center}
\end{figure}

\begin{figure*}
\begin{center}
\includegraphics{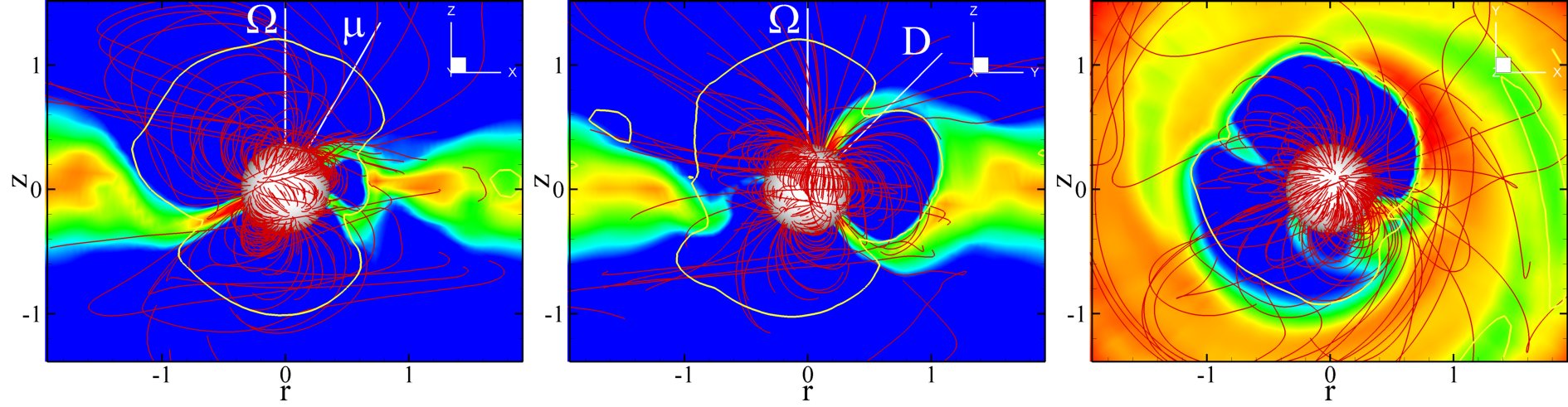}
\caption{\label{fig9} Slices of density distribution for a more general case ($\mu=0.5$, $D=0.5$, $\Theta=30^\circ$,
$\Theta_D=45^\circ$, $\Phi=90^\circ$). The left-hand, middle and right-hand panels show the projection in the $xz$, $yz$
and $xy$ planes respectively. The color background shows the density distribution varying from $\rho=0.01$ (blue) to
$\rho=2.1$ (red). The magnetic field lines are shown in red. The yellow lines shows where $\beta=1$.}
\end{center}
\end{figure*}

\begin{figure}
\begin{center}
\includegraphics{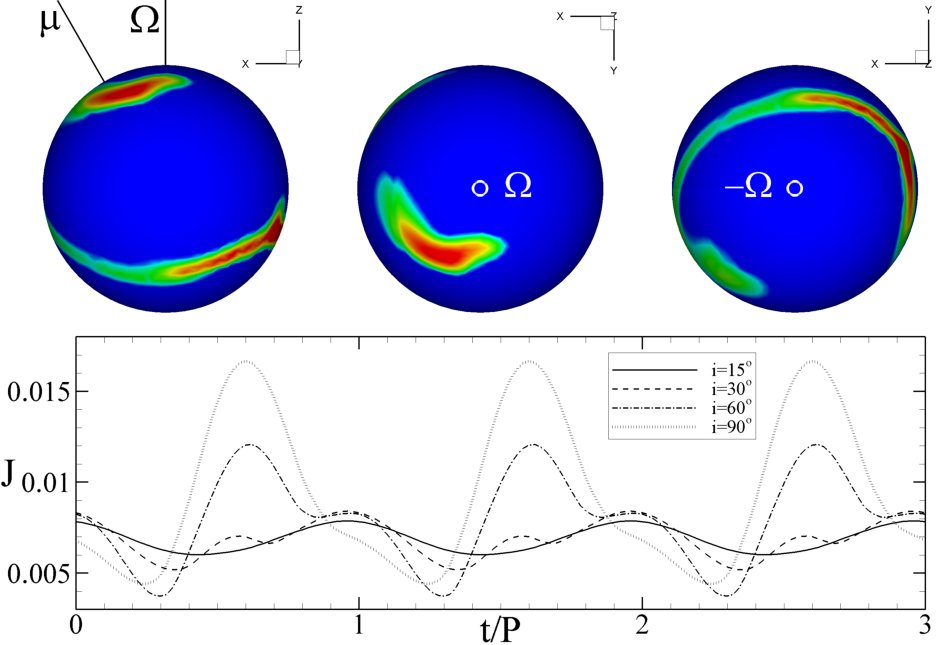}
\caption{\label{fig10} Hot spots from different angles at $t=8$ for the case of $\mu=0.5$, $D=0.5$,
$\Theta=45^\circ$, $\Theta_D=30^\circ$ and $\Phi=90^\circ$. The left-hand, middle and right-hand panels represent
edge-on, top and bottom views respectively. The red color corresponds to the densest region with maximum
density $\rho=2.3$. The bottom panel shows the light curves for different inclination angles. }
\end{center}
\end{figure}

\subsection{Off-centre dipole fields}

\begin{figure}
\begin{center}
\includegraphics{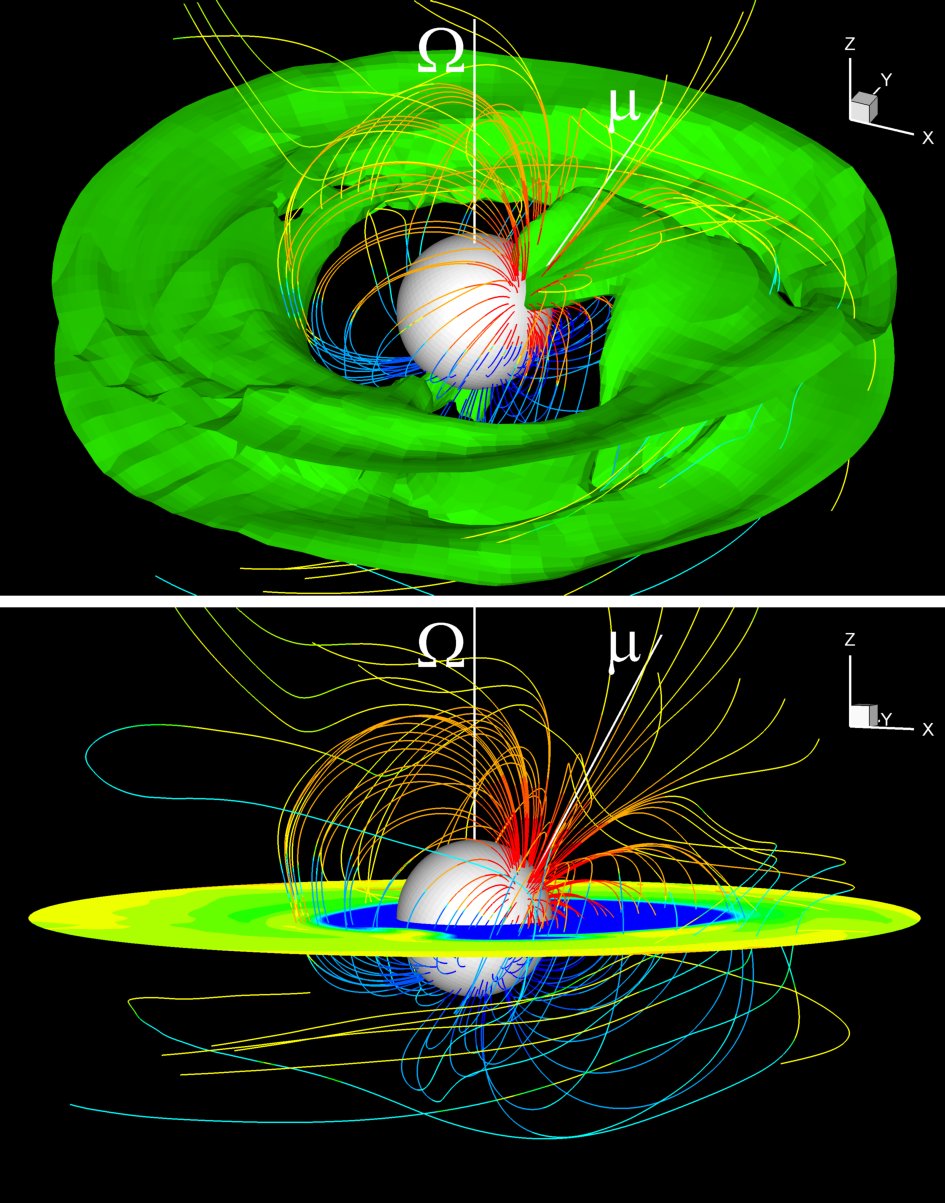}
\caption{\label{fig11} 3D views of matter flow to a star with an off-centre dipole at $t=8$, where $\mu=0.5$,
$\Theta=30^\circ$ and the dipole moment is located at $x=0.5R_*,y=0,z=0$. The green color in the top panel shows one of the density levels $\rho=0.3$. The bottom panel shows the equatorial plane with different density levels. The field lines
are shown in multiple colors.}
\end{center}
\end{figure}

\begin{figure}
\begin{center}
\includegraphics{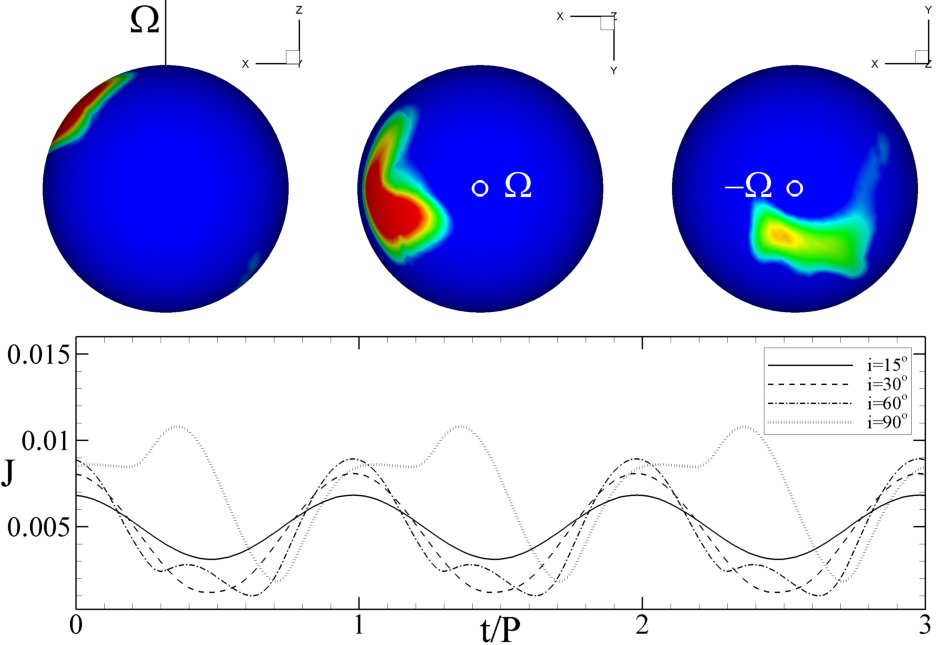}
\caption{\label{fig12} Top panel: hot spots for the off-centre case shown in Fig. 11, at $t=8$.  The red color corresponds to the densest region with maximum density $\rho=2.6$, and the yellow color in the right-hand
panel represents the density $\rho=0.7$. Bottom panel: the light curves.}
\end{center}
\end{figure}

Until now, all we discussed are cases in which the dipole and quadrupole magnetic moments are located at the center of the star. The interior convective circulations may be displaced from the center and lead to off-centre magnetic configurations. Therefore we now consider disk accretion to a star with an off-centre pure dipole field, with $\mu=0.5$. The magnetic moment is located at $x=0.5R_*$, $y=0$, $z=0$, that is, it is shifted from the centre of the star by half radius of the star. The misalignment angle is $\Theta=30^\circ$. Fig. 11 shows the 3D plot of disk accretion for this case. One can see that the north and south magnetic poles are not symmetric on the surface of the star. Due to the shift and the tilt of the dipole moment, the north magnetic pole is closer to the accretion disk than the south pole, so that most of the matter flows to the north magnetic pole. Fig. 12 shows the hot spots on the surface of the star and the corresponding light curves. The light curves are sinusoidal at $i=15^\circ,30^\circ$, and have unusual shapes at $i=60^\circ,90^\circ$.

In another set of runs, we chose a superposition of several dipoles placed at different places in the star, and with
different orientations of their axes. Fig. 13 illustrates the considered configurations of three
off-centre dipoles of equal value $\mu=0.5$, displaced by $0.4R_*$ from the center of the star in the equatorial
plane and azimuthally separated by $120^\circ$. In configuration (a), all moments are misaligned relative to the rotation
axis at $\Theta=30^\circ$, and in configuration (b), they are arranged in the equatorial plane in an anti-clockwise manner.

\begin{figure}
\begin{center}
\includegraphics[scale=0.4, angle=0]{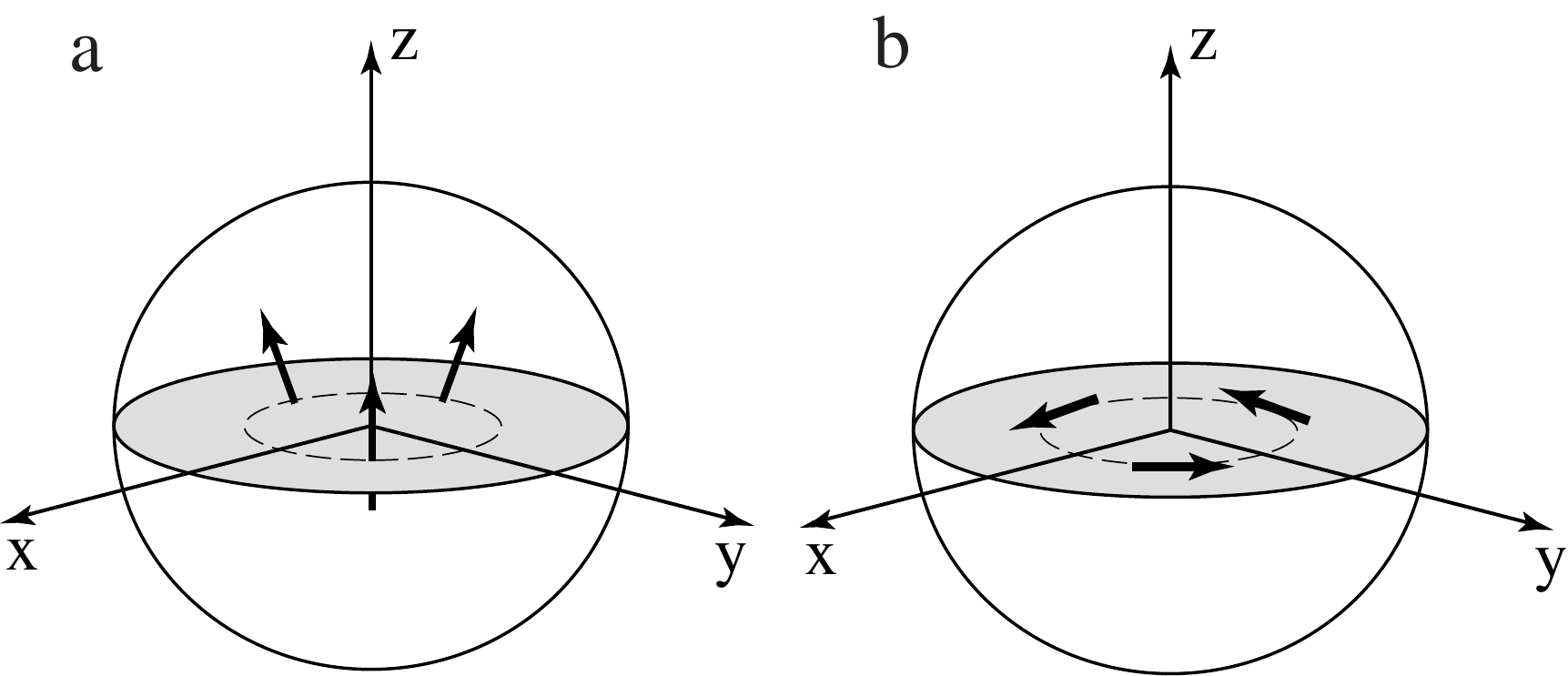}
\caption{\label{fig13} The configurations of three off-centre, equally separated dipole magnetic moments, which are
$0.4R_*$ away from the center of the star in the equatorial plane, with $\mu_1=\mu_2=\mu_3=0.5$. (a) The magnetic
moments are tilted away from the rotation axis at $\Theta=30^\circ$; (b) the magnetic moments are arranged in the equatorial plane in a circular fashion.}
\end{center}
\end{figure}

Fig. 14 shows the distribution of the magnetic field on the surface of the star. One can see that there are six poles
on the star. In case (a), three positive and three negative poles are in northern and southern hemispheres respectively. In case (b), all poles are in the equatorial planes and poles with different polarities alternate.

Fig. 15 shows a 3D view of the matter flow for the cases (a) and (b). In case (a), most of the matter flows to the north poles which are closer to the accretion disk due to the inclination of the dipoles. The magnetospheric gap is empty in the equatorial plane. In case (b), all the matter flows between loops of the closed magnetic field lines in the equatorial plane and forms some interesting equatorial funnels.

Fig. 16 shows the hot spots for the cases (a) and (b). There is a big triangular hot spot around the north pole for the case a, which spans the area near the magnetic poles. Another hot spot is present near the south pole but with very low densities. This confirms the asymmetric matter flow shown in Fig. 15. We can also see that there are no polar hot spots for case (b). Only ring-like hot spots are present in the equatorial plane, which corresponds to what we observed in Fig. 15.

\begin{figure}
\begin{center}
\includegraphics{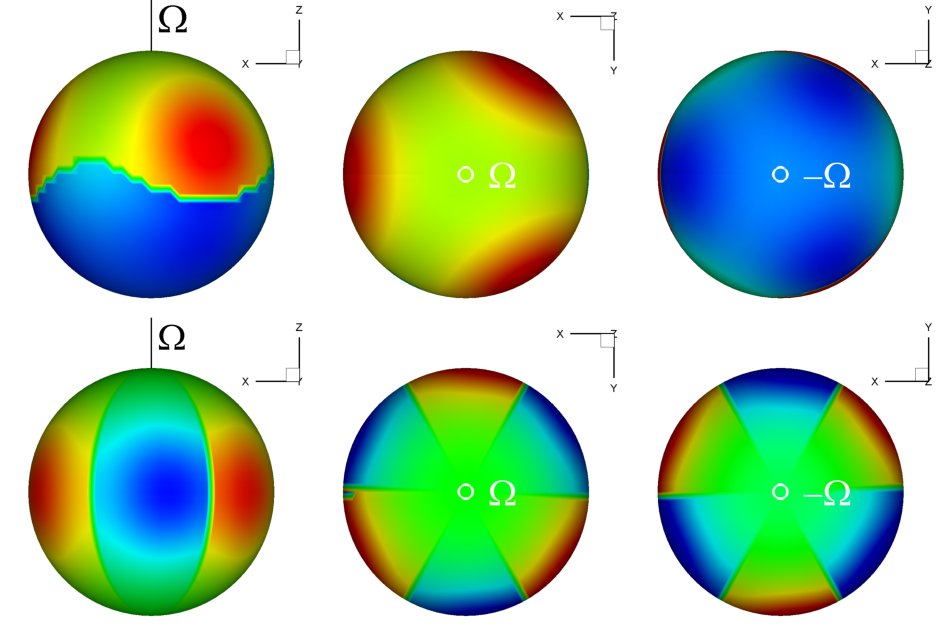}
\caption{\label{fig14} The surface magnetic field on a star with the three off-centre dipoles described in Fig. 13. Top
panel: case (a); bottom panel: case (b). }
\end{center}
\end{figure}

\begin{figure*}
\begin{center}
\includegraphics{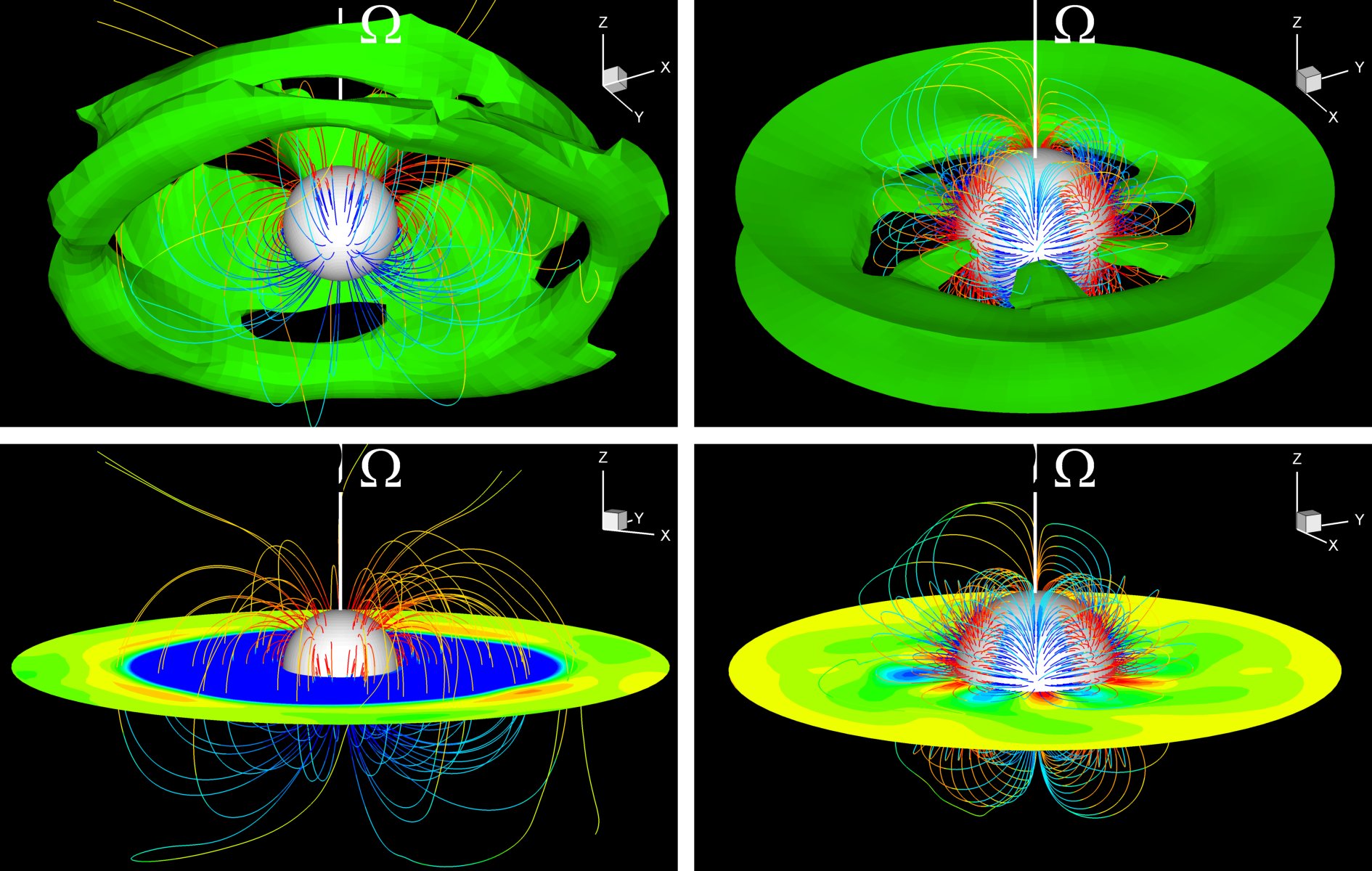}
\caption{\label{fig15} 3D views of disk accretion to a star with the three off-centered dipoles described in Fig. 13, at
$t=8$. Left-hand panel: case (a); Right-hand panel: case (b). }
\end{center}
\end{figure*}

\begin{figure}
\begin{center}
\includegraphics{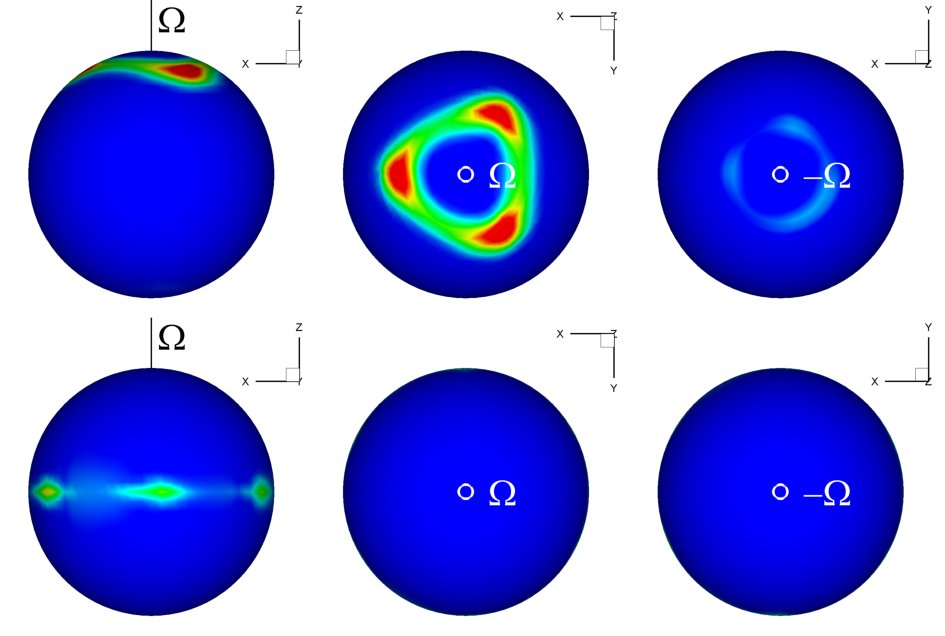}
\caption{\label{fig16} The hot spots from different angles for the cases described in Fig. 13, at $t=8$. Top panel:
case (a), the maximum densities in the hot spots in the northern and southern hemispheres are $\rho=1.4$ (red) and $\rho=0.1$ (light blue) respectively. Bottom panel: case (b), the maximum density is $\rho=0.8$. }
\end{center}
\end{figure}

\section{Contribution of the quadrupole to properties of magnetized stars}

In this section we analyze properties of magnetized stars at different ratios between the dipole and quadrupole fields.
First, we consider different configurations at the same maximum value of the field on the surface of the star. Then we
analyze different properties such as the area covered by hot spots, mass accretion rate and spin torque. Next, we fix
the dipole component but change the contribution of the quadrupole component with the main goal of understanding how strong the quadrupole should be compared with the dipole in order to change the shape of the hot spots from the pure dipole cases.

\subsection{Area covered by hot spots, and torque}

In this section we compare properties of magnetized stars of different configurations under the condition that the maximum magnetic field on the surface of the star is the the same in all cases.

We choose a number of configurations ranging from purely dipole to purely quadrupole: (a) pure dipole field, with $\mu=1.07$, $\Theta=45^\circ$; (b) pure quadrupole field, with $D=0.5$, $\Theta_D=0^\circ$; (c) aligned dipole plus quadrupole field, with $\mu=0.5$, $D=0.27$, $\Theta=\Theta_D=0^\circ$; (d) misaligned dipole plus quadrupole field, with $\mu=D=0.37$, $\Theta=0^\circ$, $\Theta_D=45^\circ$. In all cases, the maximum strength of the surface magnetic field is the same, about $B_\mathrm{max}=50$ (dimensionless value), but the contributions of the dipole and quadrupole components are different. In cases (a), (b) and (c), the field has a maximum value at the north pole, and in case (d), it is near the quadrupole axis. Here, for case (a), we choose a relatively large $\Theta$ to avoid MHD instabilities (see Kulkarni \& Romanova 2008, Romanova, Kulkarni \& Lovelace 2008). The discussed configurations are shown in Fig. 17. Below, we compare different properties of accretion to these stars.

\begin{figure}
\begin{center}
\includegraphics{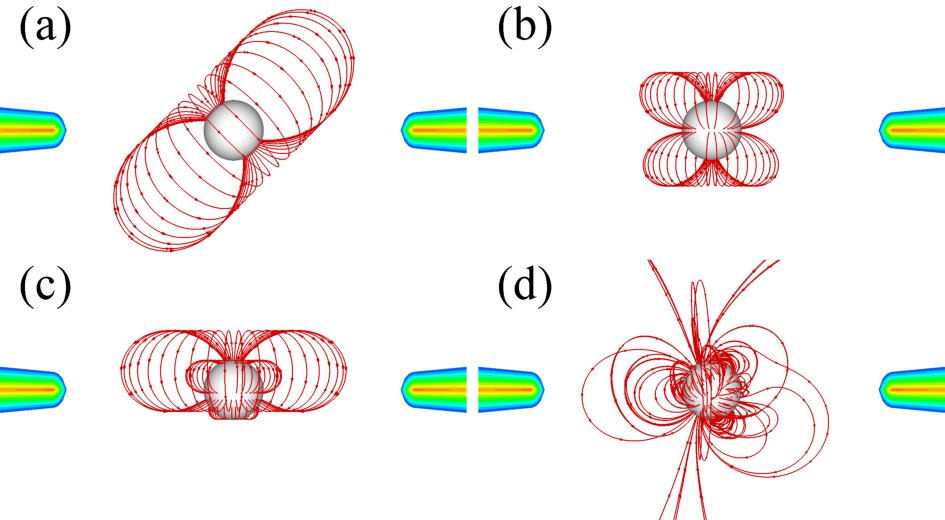}
\caption{The configurations of different magnetic fields:  case (a), pure dipole field; case (b), pure quadrupole
field; case (c), aligned dipole plus quadrupole field; case (d), misaligned dipole plus quadrupole field.}
\end{center}
\end{figure}

\subsubsection{Where does the disk stop?}

We calculated the magnetospheric radius $r_m$ in $+x$ direction in the equatorial plane and obtained the value of $r_m$
for above four cases: $r_{m,a}=1.47$, $r_{m,b}=0.62$, $r_{m,c}=1.0$, $r_{m,d}=1.05$ respectively. We can see that the
disk stops closer to the star for the pure quadrupole case than for the pure dipole case, and  $r_m$ has intermediate
values for the mixed cases. This result is expected because the quadrupole field decreases faster with distance than the
dipole field, so the the disk comes closer.

\subsubsection{Area covered by hot spots}

We also calculated the area covered by hot spots in the above cases (a)-(d). Like in Romanova et al. (2004), we chose
some density $\rho$ and calculated the area $A(\rho)$ of the region where the density is larger than $\rho$.
Then we obtain the fraction of the stellar surface covered by the spots, $f(\rho)=A(\rho)/A_*$.

We also consider the temperature distribution on the surface of the star. In spite of the possible complex processes of
radiation from the hot spots, we suggest that the total energy of the incoming stream is radiated approximately as a blackbody. The total energy flux carried by inflowing matter to the point $\mathbf{R}$ on the star is,

\begin{equation}
F_e(\mathbf{R})=\rho\mathbf{n}\cdot\mathbf{v}(\frac{1}{2}v^2+w),
\end{equation}
where $\mathbf{n}=-{\hat{\mathbf{r}}}$, $\mathbf{v}$ is the matter velocity, and $w=\gamma(p/\rho)/(\gamma-1)$ is
the specific enthalpy of the matter. So we have $F_e(\mathbf R)=\sigma T_{\mathrm{eff}}^4$ and the effective blackbody
temperature,

\begin{equation}
T_{\mathrm{eff}}=\bigg[\frac{\rho\mathbf{n}\cdot\mathbf{v}}{\sigma}(\frac{1}{2}v^2+w)\bigg]^{1/4},
\end{equation}
where $\sigma$ is the Stefan-Boltzmann constant. Similarly, we obtain the area $A(T_\mathrm{eff})$ with
effective temperature larger than $T_\mathrm{eff}$, and the fraction $f(T_\mathrm{eff})=A(T_\mathrm{eff})/A_*$.

Fig. 18 shows the distributions of $f(\rho)$ and $f(T_\mathrm{eff})$ for cases (a)-(d). We can see that the area covered by spots in the pure quadrupole case (case b) is several times smaller than that in the dipole case (case a). The area covered by spots in the mixed dipole plus quadrupole cases (b and c) is in between. We conclude that if most of the field is determined by the quadrupole, the expected area covered by spots is smaller. We do not know exactly the reason for the smaller fraction $f$ in the cases of quadrupole field. One of the factors which we should mention is that the mass accretion rate to the surface of the star is also smaller in the case of quadrupole configurations (see \S 4.1.3 and Fig. 19). It seems that it is ``harder" for matter to penetrate between the quadrupolar field lines to the poles compared with the dipole case, and matter accretes through a narrow quadrupole ``belt" (Long et al. 2007), which leads to smaller accretion rates and smaller hot spots. Fig. 18 also shows that for equal $f(\rho)$ or $f(T_\mathrm{eff})$, the density and temperature of the spots in the pure quadrupole case are smaller than in the pure dipole case.

\begin{figure}
\begin{center}
\includegraphics{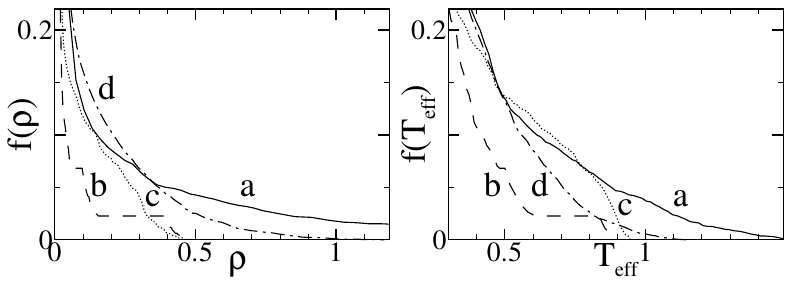}
\caption{Left-hand panel: area covered by the hot spots where the density is larger than $\rho$ (in units of
$\rho_0$, see \S2.2); right-hand panel: the area covered by the hot spots where the effective temperature is larger
than $T_{\mathrm{eff}}$ (in units of $T_{\mathrm{eff,0}}$, see \S2.2). The solid, dashed, dotted and dash-dotted lines
represent the four cases (a)-(d) shown in Fig. 17 respectively. }
\end{center}
\end{figure}

\subsubsection{Mass accretion rate and torque}

\begin{figure}
\begin{center}
\includegraphics{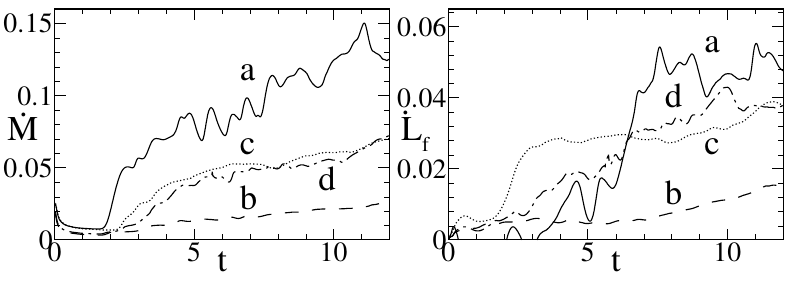}
\caption{The mass accretion rate $\dot{M}$ (left-hand panel) and the angular momentum fluxes associated with the
magnetic fields $\dot{L}_f$ (right-hand panel) for stars with different magnetic configurations. The solid, dashed,
dotted and dash-dotted lines represent the four cases (a)-(d) shown in Fig. 17 respectively. }
\end{center}
\end{figure}

As we mentioned above, matter accretion rate to the surface of the star $\dot{M}=-\int
\mathrm{d}\bm{S}\cdot\rho\bm{v}_p$ is several times smaller for the quadrupole case than for the dipole case
($\bm{v}_p$ is the poloidal component of the velocity). The mixed cases (c) and (d) have intermediate $\dot{M}$ (see
Fig. 19). We also calculated the angular momentum fluxes transported from the accretion disk and corona to stars for
cases (a)-(d). The angular momentum flux carried by the matter $\dot{L}_m=-\int\mathrm{d}\bm{S}\cdot\rho
rv_{\phi}\bm{v}_p$ is about 10 times smaller than the angular momentum flux associated with the field
$\dot{L}_f=\int\mathrm{d}\bm{S}\cdot rB_{\phi}\bm{B}_p/(4\pi)$. So we only show $\dot{L}_f$ in Fig. 19. When matter
flows in, the magnetic field lines inflate from the initial configuration. Later, they reconnect, and the transport of
angular momentum flux becomes strong. This process is longer in the pure dipole case (a). For corotation radius
$r_{cor}=2$ considered in the above cases, the star spins up mostly through the field lines connecting it to the inner
part of the disk. Here we choose slowly rotating stars ($r_{cor}=2$) for all magnetic configurations to make sure that
the magnetospheric radius $r_m$ is smaller than the corotation radius $r_{cor}$, and the star spins up in all cases. So
we can compare the positive spin torques for different magnetic geometries. Fig. 19 shows $\dot{M}$ and $\dot{L}_f$ for
cases (a)-(d). One can see that the angular momentum flux is largest in the pure dipole case (a) and smallest in the
pure quadrupole case (b). This is an interesting result because in the case of the quadrupole field, the disk comes
closer to the star and it rotates faster, so that one would expect larger angular momentum flux in the case of the
quadrupole field. But we see the opposite. We can speculate that stars with predominantly multipolar fields may not
spin up as fast as stars with a mainly dipole field. This can be explained by the smaller connectivity between closed
multipolar field lines and the inner region of the disk (see also Long et al. 2007).

\subsection{Hot spot shape}

Next, we fix the dipole magnetic field at some relatively high value $\mu=2$, the misalignment angle at some
value $\Theta=30^\circ$, and gradually increase the quadrupole component, $D=0.5,1,1.5,2$.
In all cases, the quadrupole is aligned with the $\bm\Omega$ axis. The main goal is to understand the ratio $D/\mu$ at which the properties of hot spots will depart from the dipole ones, and the ratio $D/\mu$ at which the quadrupole will strongly influence the shape of hot spots.

Fig. 20 shows that in the pure dipole case (left column), the hot spots have a typical arc-like shape. With the increase of the quadrupole component, the southern hot spot is stretched. When the quadrupole is strong enough, the southern hot spot forms a ring. We conclude that the influence of the quadrupole component on the shape of the hot spots becomes noticeable when the ratio of the quadrupole and dipole moments $D/\mu>0.25$, and becomes dominant in determining the hot spot shape when $D/\mu>0.5$. The corresponding ratios of the magnetic field strengths on the surface of the star are $B_q/B_d=0.5$ and $B_q/B_d=1$ respectively.

Fig. 21 shows the light curves for the above cases. It is interesting to note that at $i=30^\circ$ the amplitude is
relatively small for the strong quadrupole case. This can be explained by the fact that part of the southern ring-like hot spot is not visible to the observers in this orientation. When the quadrupole component become more important, the intensity and amplitude of the light curves become smaller. The shape of the light curves is more irregular than in the pure dipole configuration. Light curves are complex but may provide information about the magnetic configuration of the star, if the inclination angle is known independently.

\begin{figure}
\begin{center}
\includegraphics{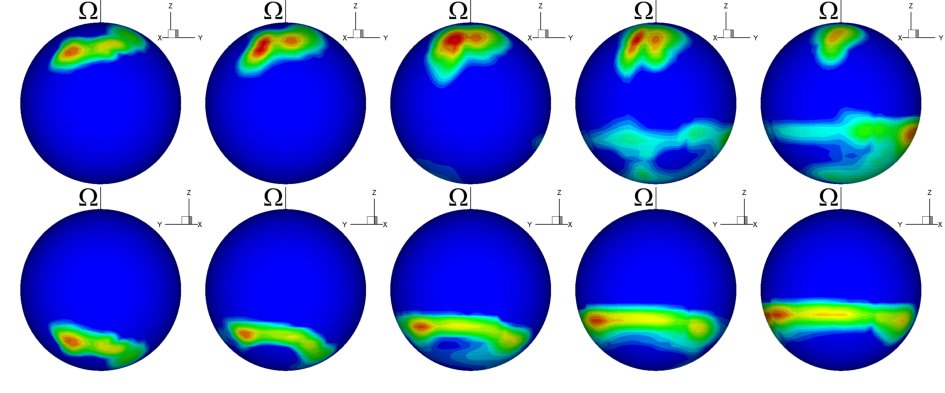}
\caption{The hot spots for different cases at $t=8$. The top row shows one side of the star, and the bottom row the
other. Each column represents different configurations $D=0$, $D=0.5$, $D=1.0$, $D=1.5$ and $D=2.0$
from the left to the right respectively.}
\end{center}
\end{figure}

\begin{figure}
\begin{center}
\includegraphics{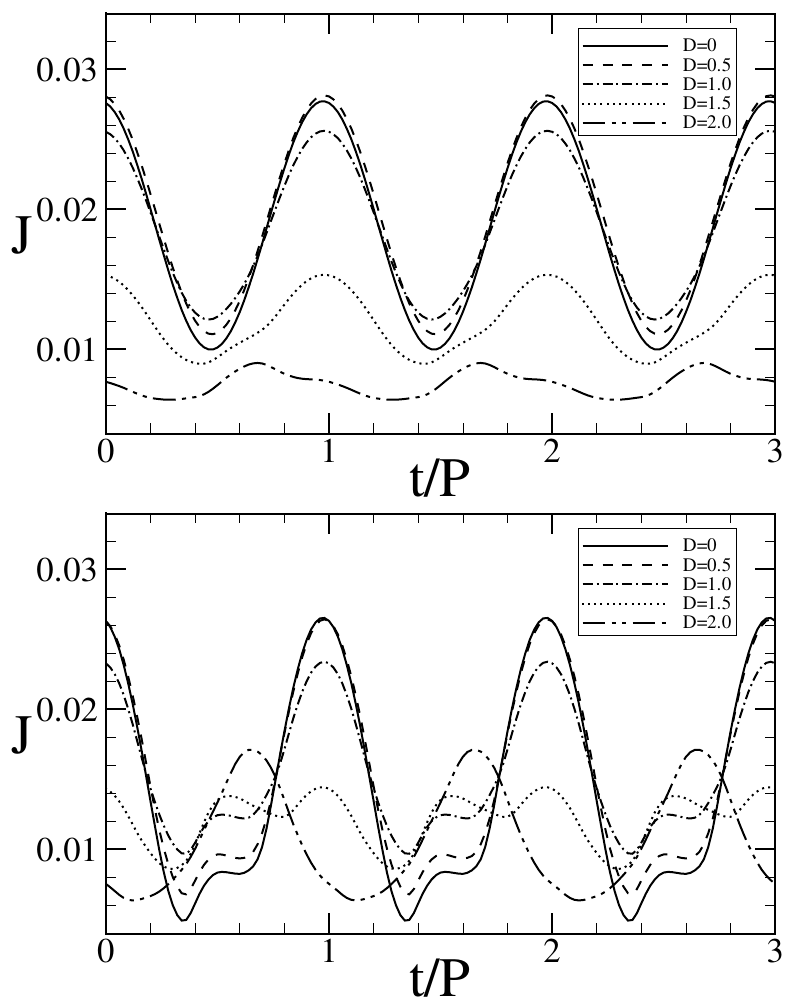}
\caption{Light curves for inclination angle $i=30^\circ$ (top panel) and $i=60^\circ$ (bottom panel) at $t=8$. The
solid, dashed, dash-dotted, dotted and dash-dot-dotted lines represent the configurations $D=0$, $D=0.5$, $D=1.0$, $D=1.5$ and $D=2.0$ respectively.}
\end{center}
\end{figure}

\section{Summary and discussion}

We investigated disk accretion to rotating stars with different complex magnetic fields using full three-dimensional
magnetohydrodynamic simulations. The main results of this work are the following:

1. We investigated accretion to a star with a dipole plus quadrupole field for general conditions where the moments
of the dipole and quadrupole components are misaligned and also where they are in different meridional planes. We
concentrated on the case where both dipole and quadrupole fields are strong so that the total field is complex. In this
case, there are three major magnetic poles with different polarities on the surface of the star, and three sets of loops
of closed magnetic field lines connecting these poles. The matter flow is not symmetric; it tends to flow to the star
along the shortest path to the nearest magnetic pole. Most of the matter flows to the star through a narrow quadrupole
``belt" between loops of field lines. Thus arc-like and ring-like hot spots typically form on the star.

2. We investigated cases with first one and then a set of off-centre dipoles. In the case of one displaced dipole,
the matter flow is not symmetric. More matter flows through one stream than the others, and one hot spot is larger than the others. In the case of three displaced dipoles, there are three positive and three negative poles on the star. Depending on the directions of the dipole moments, matter flows through multiple streams and forms strong hot spots at the north pole or in the equatorial plane.

3. The mass accretion rate, the area covered by hot spots, and the torque on the star are several times smaller in the
quadrupole-dominated case than in the dipole cases (for conditions where the maximum strength of the magnetic
field on the star's surface is the same).

4. The quadrupole component has a noticeable influence on the shape of hot spots if the magnetic field of the
quadrupole component $B_\mathrm{q}\gtrsim0.5B_\mathrm{d}$ ($D/\mu\gtrsim 0.25$). The shape of hot spots is chiefly
determined by the quadrupole field if $B_\mathrm{q}\gtrsim B_\mathrm{d}$ ($D/\mu\gtrsim0.5$).

5. The light curves from hot spots of rotating stars with complex magnetic fields are often sinusoidal in the case of
small inclination angles, $i\leqslant30^\circ$, but are more complex for $i\geqslant60^\circ$. Even for
$i\geqslant60^\circ$, the light curves often resemble those of misaligned dipoles at large $i$. However, in some cases
they are very unusual. We conclude that, (a) very unusual, non-sinusoidal light curves may be sign of the complex
fields; (b) simple sinusoidal light curves \textit{do not} rule out complex fields; (c) light curves may be a tool
for analyzing the complexity of the field, if the inclination angle is determined independently.

Many CTTSs show highly variable light curves. From our simulations we conclude that hot spots do not change their shape
or position significantly and thus may not be responsible for such strong variability. The variability might be explained by some other mechanism, such as a highly variable accretion rate, or unstable accretion (Romanova, Kulkarni \& Lovelace 2008; Kulkarni \& Romanova 2008).

Results of 3D MHD simulations of accretion to stars with complex fields can be used to compare models with
observations. For example, recently, Donati et al. (2007b) derived the magnetic topology on the surface of CTTS V2129
Oph, and found  that it has a dominant octupole component with $B_{\mathrm{oct}}\sim 1.2$kG and a weaker dipole
component with $B_d\sim0.35$kG. Most of the accretion luminosity is concentrated in a quite large high-latitude spot
($5\%$ of the total stellar surface) close to the magnetic pole. We have not modelled accretion to stars with octupole
fields, but with different contributions from dipole and quadrupole components. This gives
information on how matter flows to the star. Our analysis shows that the quadrupole dominantly determines the matter flow close to the star and the shape of hot spots if $B_q>B_d$. Thus, we would expect that in a star with a dominant octupole, the accretion spot would be determined by one of octupole belts, or part of the belt in the case of tilted octupole fields. So from our point of view, for such a strong octupole field, the single round spot in the high latitude region is an unexpected feature, which requires additional analysis in the future.

The results of our simulations are also important for analysis of magnetospheric gaps in young stars. In stars with multiple magnetic poles, for example, in the case of the dipole plus quadrupole fields with different orientations of the axes, the probability is high that one accretion stream crosses the equatorial plane, thus filling it with matter. This factor may be important in determining the survival of close-in exosolar planets, which have a peak in their spatial distribution at a few stellar radii. In the case of a dipole field, a large low-density magnetospheric gap may be formed in the equatorial plane, which may halt subsequent migration of planets (Lin et al. 1996, Romanova \& Lovelace 2006). However, if a star has a complex magnetic field with several poles, one of streams may flow through the equatorial plane and the magnetospheric gap may be not empty.

%%%%%%%%%%%%%%%%%%%%%%%%%%%%%%%%%%%%%%%%%%%%%%%%%%%%%%%%%%%%%%%%%

\section*{Acknowledgments}

This research was conducted using partly the resources of the Cornell Center for Advanced Computing, which receives
funding from Cornell University, New York State, federal agencies, foundations and corporate partners, and partly using
the NASA High End Computing Program computing systems, specifically the Columbia supercluster. The authors thank A.V.
Koldoba and G.V. Ustyugova for the earlier development of the codes, and A.K. Kulkarni for helpful discussions. This work was supported in part by NASA grants NAG 5-13060, NNG05GG77G, NNG05GL49G and by NSF grants AST-0607135, AST-0507760.

\end{document}